# Mechanophotonics: Pseudo-plastic Organic Crystal as a Fermat Spiral Optical Waveguide

Melchi Chosenyah, Avulu Vinod Kumar, and Rajadurai Chandrasekar, *Member, IEEE*

*Abstract*— An unprecedented organic Fermat spiral optical waveguide (FSOW) self-transducing green fluorescence is fabricated using a pseudo-plastic (*E*)−1-(((5-bromopyridin-2-yl)imino)methyl)naphthalene-2-ol (BPyIN) crystal. A 1.618-millimeter-long crystal is initially bent into a hairpin-like bent waveguide. Later, a meticulous mechanophotonic strategy is employed to sculpt the hairpin-like bent waveguide into the Fermat spiral geometry, covering a compact area of 330×238 μm². The optical signal in FSOW survives two sharp 180° turns to produce optical output. The remarkably low bending-induced optical loss in FSOW can be ascribed to the smooth-defect-free surface morphology of the crystal. The development of such versatile optical components capable of transducing light through sharp bends is pivotal for realizing large-scale all-organic photonic circuits.

## I. INTRODUCTION

Fermat spiral pattern has captured attention for its intrinsic beauty and governing the natural architectures like seed arrangement in sunflowers, expansion of the Milky Way galaxy, and hurricane patterns. Named after Pierre de Fermat, Fermat's spiral shown in Fig. 1, is an alluring mathematical curve that holds significance both in terms of aesthetics and practical applications. It is expressed as

$$r = \pm a\theta^{1/2} \quad (1)$$

in polar coordinates, elucidating the relationship between the radial distance (r), a constant scaling factor (a), and the angular parameter (θ). The distinctive characteristic of the Fermat spiral is that the radial separation between consecutive turns varies inversely with the angular displacement of successive turns from the spiral's center (Fig. 1) [1]. In photonics, the two termini can be utilized as light input and output points as they are pointing outside the spiral zone. Therefore, Fermat spirals are ideal candidates to delay the optical signals, and these spiral waveguides (SWs) are used for sensors, delay lines, thermo-optic phase shifters, spiral gratings, and Brillouin lasers [2-5].

In silicon photonics, the spiral arrangement of millimeter-long waveguides is precisely shaped to create extremely compact and microscopic high-density circuits [6]. However, it is difficult to produce low-loss SWs with the electron beam lithography (EBL) fabrication technique [7]. Further, the Si-based materials lack active waveguiding characteristics and are prone to breakage under mechanical stress due to high hardness

This work was supported in part by SERB New Delhi, SERB-STR/2022/00011, (Corresponding author: Rajadurai Chandrasekar.) The authors are with the School of Chemistry and Centre for Nanotechnolgy, University of Hyderabad, India 500046 (e-mail: r.chandrasekar@uohyd.ac.in).

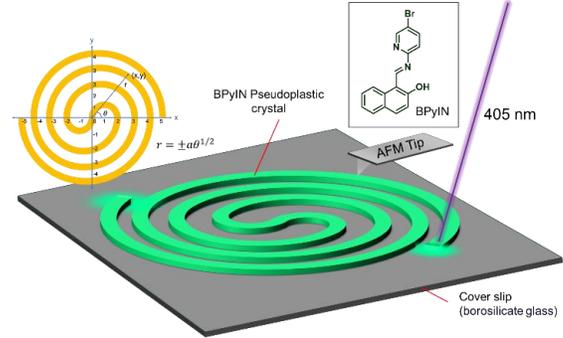

**Fig. 1.** Schematic illustration of the envisaged Fermat spiral optical waveguide (FSOW). The inset shows the molecular structure of BPyIN (right top) and Fermat spiral in cartesian coordinates (left top).

(Hardness, *H* = 180 GPa; Young's modulus = 150 GPa; and Poisson's ratio = 0.17) [8,9]. The brittleness of Si-based photonic circuits prevents post-fabrication reconfigurability, impeding their application in futuristic reconfigurable circuits. Conversely, organic crystals manifest their suitability for photonic applications as they possess supreme qualities like good solution processability, high refractive index [10,11], optical emissions (fluorescence (FL)/phosphorescence (PL)) [12,13], mechanical compliance (elastic, plastic, pseudo-plastic) [12-20], stable room-temperature exciton-polaritons [21], optical non-linearity [22], chirality [22], passive [16, 23] and active [10-23] light-guiding mechanisms, bandwidth modification using reabsorbance [14,18,19,23,24,50], and the use of evanescent field assisted energy-transfer principles [18,24,50].

The mechanically flexible crystals are mainly characterized, as either elastic (spontaneous reversible mechanical deformation) or plastic (irreversible mechanical deformation) when load is applied [11,17,23]. Recently, it has been observed that elastic microcrystals exhibit pseudo-plasticity due to their strong adherence to silica substrates [11-14,18,19,23,24]. This remarkable property of the elastic crystals combined with pseudo-plasticity, facilitated the fabrication of many proof-of-principle photonic components on silica substrate by manipulating the geometries of organic microcrystals using atomic force microscopy (AFM) tip-an innovative approach known as *mechanophotonics* [11-14,18,19,23,24]. These photonic components encompass waveguides (visible and near-IR) [11,14,16,23,25], ring-resonators [18,19,24], add-drop filters [18,19,24], modulators [26,27], directional couplers [23], wavelength division multiplexer (WDM) [28], interferometers [29], and microlasers [31]. The realization of these components underscores the promising potential of organic pseudo-plastic crystals in advancing the field of organic crystal-based photonic circuits. Recently, we demonstrated the construction of a variety of organic spiral waveguides (OSWs) with different geometries, for photonic circuit applications [50]. However, there are no reports on Fermat-type spiral architectures with organic crystals so far.

In this letter, we describe the fabrication of the first Fermat spiral optical waveguide (FSOW) based on an exceptionally flexible and green emissive crystal, (*E*)-1-(((5-bromopyridin-2-yl)imino)methyl)naphthalene-2-ol (abbreviated as BPyIN). The chemical structure of BPyIN is shown in Fig. 1. The remarkable pseudo-plasticity of BPyIN crystals on a silica substrate enables the creation of FSOWs through mechanophotonics techniques as shown in Fig. 1. by employing AFM cantilever tip micromanipulation, we transformed a millimeter-sized BPyIN crystal into a compact FSOW covering a small area (330×238 µm$^2$) that functions as a low-loss optical waveguide. These FSOWs play a pivotal role in advancing organic single-crystal interferometers, contributing to the progression of organic photonic circuit technologies.

## II. EXPERIMENTS

The fabrication of Fermat spiral shape requires a meticulous mechanophotonics strategy and the utilization of a long millimeter-sized pseudo-plastic crystal. The crystals of BPyIN, which produce green light, are a suitable choice for this investigation as shown in Fig. 2(a). In our previous study, an organic crystal optical interferometer was made using the extremely flexible BPyIN microcrystals [52]. The electronic absorption spectra of BPyIN crystals shown in Fig. 2(b) reveal optical absorption in the UV-visible region (≈300–550 nm) and the FL in the green region (≈500–650 nm; $\lambda_{max}$ ≈540 nm) of the visible spectrum. Interestingly, a significant portion of BPyIN's absorption overlapped (500 – 566 nm) with its FL which led to the reabsorption (selfabsorption). The multilayer molecular packing that is reinforced by different non-covalent interactions provides support for the remarkable mechanical flexibility of BpyIN crystals [52]. Under external stress the atoms can move from their equilibrium positions due to weak intermolecular interactions. Consequently, it is feasible for molecules to expand or contract along the extended layers on the convex and concave areas of the bent crystals. In order to prevent crystal shattering, these molecular motions dissipate the external stress. Nevertheless, the absence of stress dissipation channels causes the crystals to shatter when force is applied along other crystal planes.

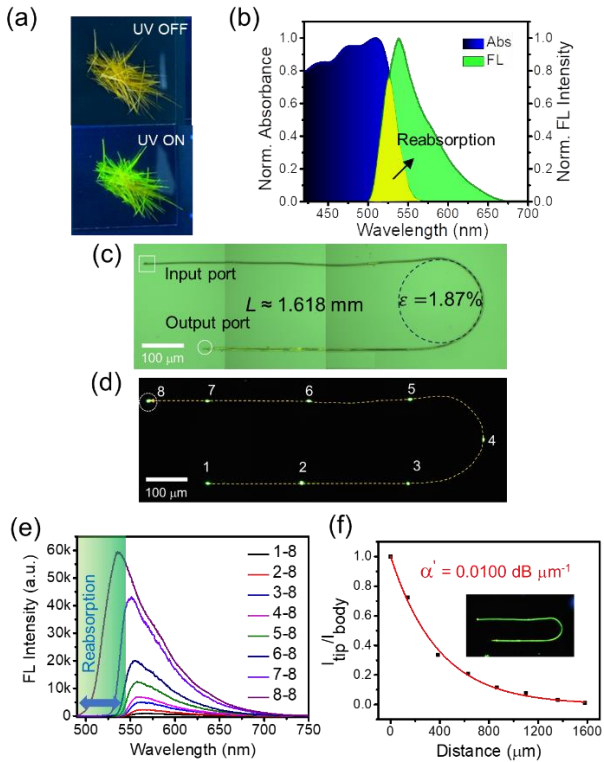

**Fig. 2.** (a) Photographs of BPyIN crystals kept on a cover slip before and after UV irradiation. (b) Photophysical properties of BPyIN in solid-state. Stitched c) confocal optical and d) laser FL microscopic image of a hairpin-like bent waveguide of the length *L*=1.618 mm. e) FL spectra collected at 8 for the different excitations from 1 to 8. e) Plot of the intensity ratio at the tip and the body of the crystal ($I_{tip}/I_{body}$) versus the optical path length, used to estimate the optical loss coefficient (α') for hairpin-like bent waveguide. (Inset: FL image of bent BPyIN)

The long bulk crystals were prepared by slow evaporation (6-7 days) of the BpyIN solution in methanol at ambient conditions. Later, a 50 µL solution of the mother liquor was drop casted onto a coverslip (borosilicate) to obtain rod-like microcrystal under a confocal microscope upon the solvent's complete evaporation.

## III. FERMAT SPIRAL OPTICAL WAVEGUIDE

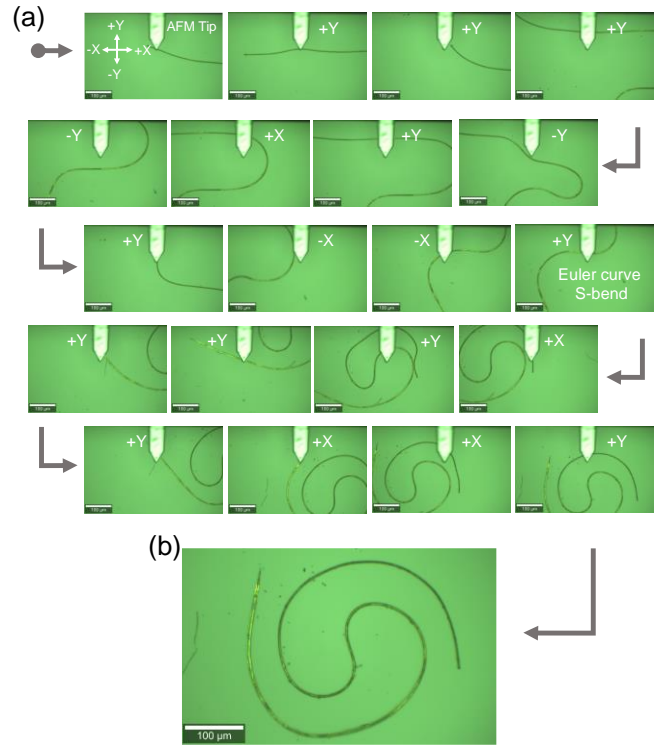

**Fig. 3.** (a) Confocal optical images showing the micromechanical manipulation of the hairpin-like bent microcrystal of BPyIN into a Fermat spiral optical waveguide (FSOW) using AFM cantilever tip. (b) Confocal optical image of fabricated FSOW. (Scale bar is 100 µm)

For the fabrication of FSOW, a BPyIN microcrystal of length ≈1.618 mm was selected and transformed into a hairpin-like bent geometry as shown in Fig. 2 (c,d). The diameter of the hairpin bend was about 200 µm. The mechanical strain stored in this hairpin bend was calculated using the below equation

$$\text{Strain } (\varepsilon) \% = t/2r \times 100 \quad (2)$$

where *t*, is the thickness of the crystal and *r* is the radius of the hypothetical circle formed around the bend. A strain of 1.87% was accumulated at this hairpin bend in the crystal. The FL self-guiding propensity of the bent BPyIN microcrystal was examined using the confocal microscope connected to a charge-coupled device spectrophotometer (CCD) detector in the transmission geometry using 60× objective to focus the continuous wave laser beam (405 nm) onto the crystal. The orthogonal illumination of the input laser beam at one of the crystal termini (input port), which is marked as a white dotted square in Fig. 2(c) produced a green FL ($\lambda_{max} \approx 540$ nm). The signal collected at the opposite terminal (output port), marked as a white dotted circle in Fig. 2(c) was collected with the 20× objective [Numerical Aperture (NA) = 0.4], reveals the generated FL is transduced along the crystal waveguide. This confirmed that the hairpin-like crystal is acting as an active optical waveguide by the outcoupled green light. Importantly, the higher energy region corresponding to ≈500-545 nm region was reabsorbed as the FL propagated to the other end due to reabsorption. The excitation-position-dependent optical waveguiding studies reveal the effective light propagation through the 180° bent crystal as shown in Fig. 2(e).

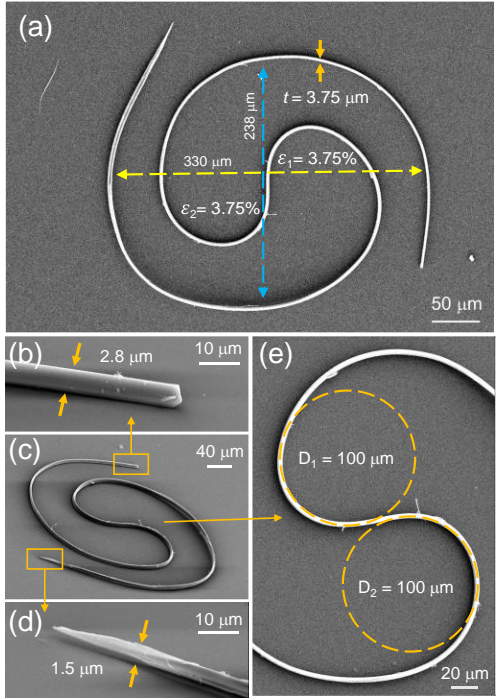

**Fig. 4.** FESEM images of fabricated FSOW (a) Top view (inset: thickness, length, and % strain values of Euler's bent S-curve. Tilt view of (b-d) fabricated FSOW and respective tips. (e) Top view of the Euler's curve S-bend.

The optical loss coefficient ($\alpha'$), a parameter describing the light transduction ability of the waveguide is calculated by the following equation

$$I_{tip}/I_{body} = e^{(-\alpha L)} \quad (3)$$

where $I_{tip}$ and $I_{body}$ correspond to FL intensities at collection and excitation positions on the waveguide. $L$ is the optical path length traveled by the photons between excitation and collection positions. The plot of intensity ratio at the tip and the body of the crystal ($I_{tip}/I_{body}$) versus the distance of the light propagation path shown in Fig. 2(f) reveals the $\alpha'$ for hairpin-like bent structure is 0.0100 dB $\mu m^{-1}$. The optical waveguiding hairpin-like crystal was subjected to manual mechanical micromanipulation using the AFM cantilever tip as shown in Fig. 3. A minimal amount of force (below the breaking threshold) was employed to deform the microcrystal into a Fermat spiral shape. The right, left, upward, and downward movements of the AFM tip are denoted as +x, -x, +y, -y, respectively. As shown in Fig 3(a,b), initially, by using the bending micromechanical operation [13], the hairpin-like bent crystal is meticulously carved into Euler's type curve, S-bend in the middle, forming the foundation of the Fermat spiral.

Subsequent sequential maneuvering of the remaining portions of the crystal leads to the attainment of the desired Fermat spiral shape as shown in Fig. 3(b). The compactly packed Fermat spiral-shaped crystal sustained high vacuum electron beam imaging conditions, illustrating the stability of the fabricated structure. The FESEM study shown in Fig. 4 shows the smooth surface of the waveguide without any visible defects. The horizontal and vertical lengths of the fabricated structure corresponded to 330 and 238 µm, respectively as shown in Fig. 4(a). The thickness of Fermat spiral-shaped crystal all along the spiral is almost the same ($t$ =3.75 µm) except for the two termini regions of the crystal. Fig. 4(b-d) shows the thickness of the top and bottom facet termini of about 1.5 µm and 2.8 µm, respectively. A hypothetical circle is superimposed on the upward and downward-curved portions of the Euler's curve in Fermat spiral-shaped crystal to estimate the strain stored in these two bent regions using equation 2, which are $\varepsilon_1$=3.75%, and $\varepsilon_2$=3.75%, respectively as shown in Fig. 4(e).

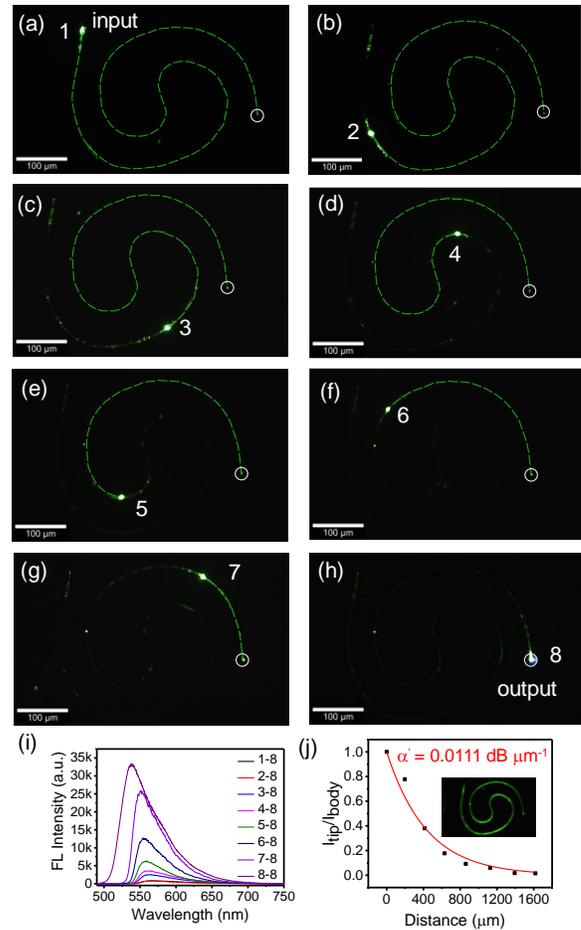

**Fig. 5.** (a-h) FL images of the excitation position-dependent waveguiding experiment of the FSOW. (Dashed green lines – light propagation path, Scale bar is 100 µm) (i) corresponding FL spectra. (j) Plot of the intensity ratio at the tip and the body of the crystal ($I_{tip}/I_{body}$) versus the optical path length, used to estimate the optical loss coefficient ($\alpha'$) for FSOW. (Inset: FL image of the fabricated FSOW).

To understand the waveguiding propensity of the fabricated FSOW, the left tip of the FSOW (labeled as input) in Fig. 5(a) was optically excited with the 405 nm solid-state laser. At the excitation point, the FSOW produced a green FL, which propagates along the waveguide's long curved axis towards the other terminal of the FSOW. The output was collected at the right tip of the FSOW tip (labeled as output; dotted white circle). As the green FL traveled along the FSOW, it survived two 180° turns in order to reflect at the other terminal of FSOW. Such versatile optical waveguides are essential for constructing multi-component large-scale organic photonic circuits.

The FSOW was optically excited at different positions from regions 1-8 as shown in Fig 5(a-h), and the output spectra were collected at the right bottom tip and these excitation-position dependent FL spectra were utilized to estimate the $α'$ of FSOW using equation 3 to be 0.0111 dB $μm^{-1}$ (Fig. 5b,c). Surprisingly, the optical loss in the FSOW was remarkably low. The very minimal bending-induced optical loss, $α' = α'_{Fermat\ spiral} - α'_{Hairpin-like\ bent}$ is 0.001 dB $μm^{-1}$ can be ascribed to the defect-free high-quality crystals of BPyIN. The demonstration of such versatile light-guiding optical components is central to the development of functional organic photonic circuits.

IV. CONCLUSION

In summary, we have successfully created an organic single-crystal Fermat spiral optical waveguide (FSOW) using the innovative mechanophotonics technique. This involved employing highly flexible and green emitting crystal of (*E*)−1-(((5-bromopyridin-2-yl)imino)methyl)naphthalene-2-ol (BPyIN). The millimeter-long microcrystal was intricately shaped into a hairpin-like structure, and subsequent waveguiding studies were conducted. The clever use of the microcrystal's pseudo-plasticity played a key role in achieving the FSOW. The remarkably low optical loss induced by bending underscores the effective light-guiding capabilities of the fabricated organic single crystal FSOW. By emulating such photonic components through the seamless integration of pseudo-plasticity and mechanophotonics, we affirm the superiority of organic crystals and lay the groundwork for realizing large-scale all-organic single crystal photonic circuits in the near future.